\documentstyle[12pt,amssymb,latexsym,amsmath]{article}

\newcommand{\cosech}{\mathop{\mathrm{cosech}}\nolimits}

\begin{document}
\title{ Fourier form of the dressing method: simple example of
integrable (2+1)-dimensional integral-differential equation 
} 

\author{ Alexandre I. Zenchuk\\
Center of Nonlinear Studies of\\ L.D.Landau Institute 
for Theoretical Physics  \\
(International Institute of Nonlinear Science)\\
Kosygina 2, Moscow, Russia 119334\\
E-mail: zenchuk@itp.ac.ru\\}
\maketitle

\begin{abstract}
 We represent the Fourier form of the dressing method, which is
 effective for construction of multidimensional
 integral-differential equations together with their solutions.
 Example of integrable (but non-physical) 
 expansion  of Intermediate Long Wave
 equation  in (2+1)-dimensions  is considered.
\end{abstract}

\section{Introduction}

Dressing method represents a productive
tool for solving large family of nonlinear equations of
Mathematical Physics.
Many equations from this family have well known applications in
different branches of mathematical Physics: hydrodynamics, plasma
physics, optics, superconductivity. 
We mention three  classical versions of this method as a simple
way for construction of  particular solutions to those nonlinear
Partial Differential
  Equations (PDE)
which are integrable by the Inverse Spectral Transform (IST) (so
called completely integrable PDE).
First one is Zakharov-Shabat dressing method \cite{ZSh1,ZSh2},
which uses integral      Volterra operators. In particular, 
it solves initial value problem. Second version,
$\bar{\partial}$-dressing method \cite{ZM,BM,K}, is based on the
Fredholm operators.
  This algorithm is simpler in
 comparison with Zakharov-Shabat dressing method,
  however, the
 manifold of  particular solutions   is more restricted.
  Different properties of integrable PDE 
have mostly evident
interpretation  in
terms of this method (Miura and B\"aclund transformations,
commuting flows, relations among different hierarchies of
equations). Finally, one should refer to Sato 
approach to the integrability \cite{OSTT}, 
which is based on properties of  pseudo-differential operators. 
Two last versions have been used for  development of the new
version of the 
dressing method,  applicable to wide class of  nonlinear PDE, 
which are not  integrable by 
 IST \cite{Z_Sato,Z_Nw}. Some applications of these equations has
 been outlined.
 
 It is interesting to note that this new version
of the dressing method  can be readily written in Fourier form,
Sec.2. 
 Advantage of
this form is evident in work with integral-differential equations.
Below we consider an example leading to extension of 
Intermediate Long Wave equation (ILW) into  (2+1)-dimensions:
\begin{eqnarray}\label{Tx}
 v_t-  T v_{x_1x_1} -   (v^2)_{x_1} =0 ,\;\;
 Tv(x_1,x_2)\equiv \frac{\pi}{ \delta}
 \int_{-\infty}^\infty
 \coth\left(\frac{\pi}{2\delta}(\xi-x_2)\right) v(x_1,\xi) d\xi =0.
\end{eqnarray}
Reduction $\partial_{x_2}=\partial_{x_1}$ transforms this equation into 
(1+1)-dimensional 
 ILW \cite{SAK,KAS,LR,S}. We provide overdetermined
 linear system for the eq. (\ref{Tk}). 
 Then we give the algorithm for
  construction of solutions (Sec.3) and derive formular for soliton.   
 Then we consider limits of  large  and small $\delta$ (Sec.4),
 and figure out  the asymptotic integrability in the later case
 \cite{KM}.  Finally we  give
 some remarks (Sec.5). 

Note, that eq.(\ref{Tx}) is integrable in classical sence, since
it is compatibility condition for some overdetermined
 linear system. We don't consider other types of equations in
 this paper, which is aimed  on the description of the
 algorithm for
 the dressing procedure in Fourier form.
  
\section{Dressing method}

We consider two-dimensional $x$-space, $x=(x_1,x_2)$, and
time, $t$.
Real variables $x_1$ and $x_2$ are related with complex spectral parameters
$k_1$ and $k_2$, forming two-dimensional  spectral space 
$k=(k_1,k_2)$ due to Fourier transform
\begin{eqnarray}\label{F}
f(x)=\int\limits_{D_{k}}
f(k) e^{i (k_1 x_1 + k_2 x_2)} dk_1
dk_2,
\end{eqnarray}
where $D_{k}$ is 2-dimensional, either complex or real  
 $(k_1,k_2)$-space, and
$f(k) \to 0$ as $k\to\infty$.
Thus this transformation maps space of functions of two complex
variables into the space of functions of two real variables. This
means that inverse transform may not be unique. This fact will
not been discussed hereafter since inverse transform will not been used
in this paper. 

Our algorithm is based on the following integral equation 
\begin{eqnarray}\label{basic}
\Phi(k)\equiv \Phi(\lambda,\mu;k)=
\int\limits_{D_\nu}d\nu \int\limits_{D_q}\Psi(\lambda,\nu;k-q) 
U(\nu,\mu;q)  dq\equiv \Psi(k)*U(k),
\end{eqnarray}
which should be uniquely solvable for $U$: $U=\Psi^{-1}*\Phi$.
Here we use another type of parameters denoted by Greek. In
general these parameters are vector \cite{Z_Nw} but in our
consideration they are just complex parameters; 
$D_\nu$ is complex $\nu$-plane.
We use $*$ to abbreviate  combination of two types of integrals:
integral with respect to "inside" Greek parameter ($\nu$ in
eq.(\ref{basic})) and over space
of spectral parameters ($q=(q_1,q_2)$ in
eq.(\ref{basic})). In general
\begin{eqnarray}
f(k)*g(k)=\int\limits_{D_\nu} d\nu\int\limits_{D_q}
f(\mu_1,\dots,\nu;k-q) g(\nu,\dots;q) dq.
\end{eqnarray} 
 To avoid unambiguity  we 
will write explicitly dependence on $k$ while omit dependence
on Greek parameters from notations in most of the cases. 

Of principal meaning is  how to introduce the dependence on 
parameters $x$ (or $k$) and $t$. This defines particular nonlinear
system of PDE which  will be derived in result. 

Let
\begin{eqnarray}\label{var1}
k_1 \Psi(k) &=& \Phi(k) *c(k) + k_1\delta(k),\\\label{var2}
e^{-2\delta k_2}\Psi(k) &=& \Phi(k)* \left(c(k) e^{-2\delta
k_2}\right) + 
\Psi(k) + (e^{-2\delta k_2}-1) \delta(k),\\\label{var3}
\Psi_t(k)&=&i(k_1\Phi(k))*c(k) - i\Phi(k)*(k_1 c(k)),\\\nonumber
c(\lambda,\nu;k)&= &c_1(\lambda) c_2(\nu;k).
\end{eqnarray}
Being  overdetermined system of equations for the
function $\Psi$, it should be compatible, which leads to
additional conditions  for the functions $\Phi$ and $c_2$. To derive
them let us multiply eqs. (\ref{var1}) and  (\ref{var2}) 
by $e^{-2\delta k_2}$ and $k_1$ respectively and subtract one from
another. The result is splitted into two equations:
\begin{eqnarray}\label{comp1}
\left(\left(k_1  -e^{-2 \delta k_2} \right)\Phi (k)\right)*
\left(c(k) e^{-2 \delta k_2} \right)
=0,\;\;\\\label{comp12}
\Phi(k)*\left((k_1 e^{-2\delta k_2}+1) c_2\right)=0.
\end{eqnarray}
Similarly, differentiating (\ref{var1}) with respect to $t$ and
using 
(\ref{var3}) we get two other conditions
\begin{eqnarray}\label{comp2}
\Phi_t(k)=ik_1^2 \Phi(k),\;\;\\\label{comp22}
{c_2}_t(k)=-ik_1^2 c_2(k) .
\end{eqnarray}
Eqs. (\ref{comp1}-\ref{comp22}) suggest the following form
for  
  $\Phi$ and $c_2 $:
\begin{eqnarray}\label{Phi}
\Phi(k)= \Phi_0(k_1)e^{ik^2_1 t}\delta\left(e^{-2 \delta k_2}-k_1\right),\\
c_2(k)=c_0(k_1) e^{-ik^2_1 t} \delta\left(e^{2 \delta k_2}+k_1\right),\\
\end{eqnarray}
where $\Phi_0(k)$ and $c_0(k)$ are arbitrary functions of argument.
From another point of view equations
(\ref{comp1},\ref{comp2}) lead to nonlinear equation for $U$. In
fact, due to (\ref{Phi})  we can rewrite eq.(\ref{comp1}) in the form
\begin{eqnarray}\label{comp1_2}
\left(k_1  -e^{-2 \delta k_2} \right)\Phi (k) = 
\Phi_0(k_1)e^{ik_1^2 t}
\left(k_1  -e^{-2 \delta k_2} \right)
\delta\left(e^{-2 \delta k_2} -k_1  \right)\equiv F(k),
\end{eqnarray}
thus $g(k)*F(k)\equiv 0$ for any well defined function $g$.
 Now, let us substitute (\ref{basic}) for $\Phi$ in 
 eqs.~(\ref{comp1_2},\ref{comp2}), use eqs. (\ref{var1}-\ref{var3}) and
 apply operator $\Psi^{-1}(k)$
from the left.  This results in
\begin{eqnarray}\label{lin1}
 U(k)*(c_1 c_2(k)) *U(k) + k_1 U(k) -
U(k)*\left(c_1 c_2(k) e^{-2 \delta k_2}\right) *
(e^{-2 \delta k_2} U(k))-
\\\nonumber
e^{-2 \delta k_2} U(k)=0,\\\label{lin2}
 -i U_t(k) -2 U(k)*(k_1 c_1)*U(k)-
2 U(k)* c_1*(k_1 U(k)) -k_1^2 U(k)=0.
\end{eqnarray}
Introduce variables
\begin{eqnarray}
w(k)=c_{2}(k)*U(k)*c_{1},\;\;V(k)=c_2(k)*(k_1 U(k))*c_1,
\end{eqnarray}
where 
\begin{eqnarray}
f(k)*g=\int\limits_{D_\nu} 
f(\mu_1,\dots,\nu;k) g(\nu,\dots) d\nu.
\end{eqnarray}
Apply operators $c_{2}(k)$ and $c_{1}$ from the left and from the
right respectively to the eqs.(\ref{lin1},\ref{lin2}). 
 We get
 \begin{eqnarray}
V(k) (1-e^{-2 \delta k_2}) =w(k)*((e^{-2 \delta k_2} -1) w(k)) -
(k_1 e^{-2 \delta k_2}) w(k),\\
-i{w(k)}_t -2 k_1 V(k) + k_1^2 w(k) -2 w(k)*(k_1 w(k)) = 0.
\end{eqnarray}
Deriving the first of the above equations  
we used the fact that $c_2(k)\sim \delta\left(e^{2 \delta
k_2} + k_1 \right)$ which  identifies $c_1 e^{2 \delta k_2}$ with
$-c_1 k_1$ in this context.
Eliminating $V$ and introducing 
$u={w}{ (e^{-2 \delta k_2}-1)}$ 
we end up with equation 
\begin{eqnarray}\label{Tk}
i u_t-k_1^2 \coth(k_2 \delta) u +
k_1 \int\limits_{D_q} u(k-q) u(q) dq =0 .
\end{eqnarray}

This equation can be considered as compatibility condition of
appropriate linear overdetermined system, which may be readily
derived. In fact, apply the operator
$c_1$  to the eqs. (\ref{lin1},\ref{lin2}) from the right side.
Resulting equations may be written in terms of  the "spectral"
function $\chi(k)= U(k)*c_1$: 
\begin{eqnarray}\label{lin21}
k_1 \chi(k) + \chi(k)* \left((1-e^{-2\delta k_2}) w(k)\right)-
e^{-2\delta k_2}
\chi(k)=0,\\\label{lin22}
i \chi_t(k) =-2 \chi(k)* (k_1 w(k)) - k_1^2 \chi(k) .
\end{eqnarray}
This is linear system which we need. To check this  statement,
let us differentiate (\ref{lin21}) with respect to $t$
and use (\ref{lin22}). We
result in  (\ref{Tk}).

Transformation of the derived equations 
into  $x$-form is straightforward if $k_1$
and $k_2$ are real and (\ref{F})
is usual Fourier transform. Then $x$-form of $\coth(k_2 \delta)$
is $\frac{-i\pi}{\delta} \coth\left(\frac{ \pi x_2}{2
\delta}\right)$
and  we receive eq. (\ref{Tx}). In general case of complex $k$
equation (\ref{Tk}) is better to multiply by 
$\left(1-e^{-2 \delta k_2}\right)$. Then we may formally write it
in $x$-form using shift operator
${\cal{D}}f(x_1,x_2)=f(x_1,x_2+2 i \delta)$:
\begin{eqnarray}\label{D}
(1-{\cal{D}})\left(i u_t - (u^2)_{x_1}\right) + 
(1+{\cal{D}}) u_{x_1x_1}=0.
\end{eqnarray}

We can  write the above linear system (\ref{lin21},\ref{lin22}) 
in $x$-form using
transformation (\ref{F}):
\begin{eqnarray}
&&i  \chi^+_{x_1} + \chi^+u + \chi^{-}=0,\;\;
\chi^-(x_1,x_2)=\chi^+(x_1,x_2 +  2 i
\delta),\\
&&i\chi^+_t -2 i \chi^+ w_{x_1} -\chi^+_{x_1x_1}=0,\\\nonumber
&&u(x_1,x_2)=w(x_1,x_2+2 i \delta)-w(x_1,x_2),\;\;w=-u - i T u.
\end{eqnarray}
After reduction $\partial_y=\partial_x$ 
 this system becomes linear overdetermined system  for ILW \cite{KAS}.
 
\section{Solutions}

Find $\Psi$ from the eq.(\ref{var1}) and substitute it into the 
eq.(\ref{basic}). Thus we receive
\begin{eqnarray}\label{u}
\Phi(k) = \tilde \Psi(k) *U(k)+ U(k),\;\;\tilde
\Psi(k)=\frac{1}{k_1} (\Phi(k)*c(k)),
\end{eqnarray}
where
\begin{eqnarray}
\Phi(k)\equiv \Phi(\lambda,\mu;k)=\Phi_0(\lambda,\mu;k_1) e^{ik_1^2
t} \delta\left( e^{-2\delta k_2} -k_1\right),\\
c(k)\equiv c_1(\lambda) c_2(\mu;k) ,\;\;
c_2(k)\equiv c_2(\mu;k)=c_0(\mu;k_1) e^{-ik_1^2
t} \delta\left( e^{2\delta k_2} +k_1\right).
\end{eqnarray}
Thus
\begin{eqnarray}
\tilde \Psi(k)\equiv \tilde\Psi(\lambda,\mu;k) =
\int\limits_{D_\nu}\frac{d\nu}{k_1}
\left(\Phi_0(\lambda,\nu;q) 
e^{i q^2 t}\right)\Big|_{q=\frac{-k_1
e^{-2 \delta k_2}}{1-e^{-2 \delta k_2}}}
\left(c_1(\nu) c_2(\mu;q)e^{-iq^2 t}\right)
\Big|_{q=\frac{k_1 }{1-e^{-2 \delta k_2}} } 
\end{eqnarray}
Hereafter it is convenient to represent the above equation
(\ref{u}) in
$x$-form:
\begin{eqnarray}\label{ux}
\Phi(\lambda,\mu;x) = \int\limits_{D_\nu}
\tilde \Psi(\lambda,\nu;x) U(\nu,\mu;x) d\nu +
U(\lambda,\mu;x).
\end{eqnarray}

Solving (\ref{ux}) for $U$ we calculate 
$w(x)=\int\limits_{D_\nu}\int\limits_{D_\mu} c_{2}(\mu;x)  U(\mu,\nu;x)
c_{1}(\nu)$
and $u(x)=  w(x_1,x_2 + 2 \delta i)-w(x)$.
Note, that equation (\ref{ux}) 
can be solved exactly in particular case of
degenerate functions
$\tilde
\Psi(\lambda,\mu;x)=\sum_j\psi_{1j}(\lambda;x)\psi_{2j}(\mu;x) $.

In the simplest case let $c_1(\lambda)=\delta(\lambda)$, 
$\Phi_0(\lambda,\mu;k_1)=r_1(\lambda,\mu)\delta(k_1-a)$,
$c_0(\mu;k_1)=r_2 \delta(\mu)\delta(k_1-b)$.
Then one can easy receive
\begin{eqnarray}
\Phi(\lambda,\mu;x) = \tilde\Psi(\lambda,0;x) U(0,\mu;x) + 
U(\lambda,\mu;x),
\end{eqnarray}
where 
$\tilde\Psi(\lambda,0;x)$ is $x$-form of the function
$$\tilde\Psi(\lambda,0;k)=
\frac{1}{k_1} \Phi_{0}(\lambda,0;q)e^{i q^2 t}\Big|_{q=\frac{-k_1
e^{-2 \delta k_2}}{1-e^{-2 \delta k_2}}} r_2 \delta(q-b) e^{-iq^2
t}\Big|_{q=\frac{k_1 }{1-e^{-2 \delta k_2}} } .
$$
One has
\begin{eqnarray}
U(0,\mu;x)=\frac{\Phi(0,\mu;x)}{\tilde \Psi(0,0;x) + 1}.
\end{eqnarray}
Then we get line soliton solution of (\ref{Tk})
($
r_1(0,0)= r_1 
$):
\begin{eqnarray}
u&=&\frac{1}{R_1 + R_2 \cosh(\eta)},\\
\nonumber
&&R_1=\frac{\alpha}{2\beta^2},\;\;
R_2=\frac{\sqrt{\alpha^2+\beta^2}}{2\beta^2},\\\nonumber
&&\eta=- 2\beta t_1 +
\frac{\arctan(\beta/\alpha)}{\delta} t_2+ 4\alpha \beta t ,
\\\nonumber
&&a=\alpha+ i\beta,\;\;b=-\alpha + i \beta,\;\;r_1 r_2=  2 i
\beta e^{- i\arctan(\beta/\alpha)}  .
\end{eqnarray}

\section{Limits $\delta\to \infty$ and $\delta \to 0$}

As we mansion above, the reduction $\partial_x=\partial_y$ transforms 
eq.(\ref{Tx}) into (1+1)-dimensional ILW, which in turn
has two reductions leading to appropriate completely
integrable systems: $\delta\to \infty$ (Benjamin-Ono equation
(BO)
\cite{B,O}) and
$\delta\to 0$ (Korteweg-de Vries equation  (KdV) \cite{KV}).   
We consider these reductions in application to our
(2+1)-dimensional equation. We will see that limit
$\delta\to\infty$ results in another completely integrable system
with Hilbert operator instead of $T$-operator. But limit
$\delta\to 0$ does not lead to completely integrable system.
Instead we get the system, which  is integrable in "asymptotic" 
sence \cite{KM}.

First, we rewrite this  system in terms of functions
$\psi^{\pm}$, defined by  $\chi^+=e^{i (f x_1 + g x_2 +h t)}\psi^+$,
 and use variable $\tau$ instead of $t$: $
 \partial_t=\partial_\tau + \alpha \partial_{x_1}$. One has
\begin{eqnarray}\label{linsyst1}
i \psi^+_{x_1} + \psi^+ (u-\lambda) =\mu \psi^-,\\\label{linsyst2}
i\psi_\tau^+ - 2 i (\lambda +1/(2\delta)) \psi^+_{x_1} -
\psi^+_{x_1x_1}
- (2 i  w_{x_1}+\nu)\psi^+ =0,\\\nonumber
\alpha=-\frac{1}{\delta},\;\;h=\lambda^2 +\nu,\;\;
f=\lambda=-k \coth(2k \delta),\;\;e^{-2 g\delta} =-\mu
=-k\cosech(2 k\delta),
\end{eqnarray}

1. Let $\delta \to \infty$. Then operator $T$ in (\ref{Tx})
transforms into Hilbert operator
$$
T(f) \simeq 2 \int\limits_{-\infty}^\infty \frac{d\xi}{\xi-x_2}
f(\xi)
$$
and one should take $\mu=2 k$, $\lambda= -k $ in
eq.(\ref{linsyst1},\ref{linsyst2}). Functions $\psi^\pm$ are
analytical in upper and lower half of complex $z$ plane: $Re(z)=x$.
If $\partial_{x_1}=\partial_{x_2}$, then our nonlinear equation
 becomes BO.

2. Consider the limit $\delta\to 0$. Let $u=\delta U$, $\delta\ll
1$,   and expand the above linear
system in the series up to $\delta^2$.
One has:
\begin{eqnarray}
i (\psi_x-\psi_y) + \delta (\psi_{yy} + \psi U + k^2 \psi) +
\frac{2i\delta^2}{3}\left(\psi_{yyy} + k^2 \psi_y \right)&=&0,\\
i \psi_\tau - \psi_{yy} - \psi ( 2 i w-\nu)  - {i \delta}
\left( 2 \psi_{yyy}+
\frac{2}{3}\psi_y (3 U+ k^2) + \psi(U_x+U_y)
\right)+&&\\\nonumber
\frac{\delta^2}{3} \left( 
7 \psi_{yyyy} +6\psi_{yy} (U+k^2) +6 \psi_y U_y+\psi (3 U_{yy}
+3 U^2 +2 k^2 U - k^4) 
\right)&=&0
\end{eqnarray}
This system is compatible up to $\delta^2$ if
\begin{eqnarray}\label{lim2}
\delta v_t - v_x +\partial^{-1}_y v_{xx} -\frac{\delta^2}{3}
v_{xyy} + 4 \delta^2 v v_x=0,\;\; v=  w_y,\;\;
U= 2 i (w_y + \delta i w_{yy}).
\end{eqnarray}
We conclude that  derived equation (\ref{lim2})  provides 
compatibility of related overdetermined system 
only up to the order $\delta^2$.  Thus this
equation can be treated as   "asymptotically"  integrable
\cite{KM}. 

\section{Remarks}

Although we have considered only completely integrable example,
Fourier form of the dressing method can be  
 expanded  on  other types of   systems, discussed in
\cite{Z_Nw}.  Their physical application is one of the mostly
interesting problems now.   
One should note that similar "symbolic" approach to nonlinear
PDE has been used by other authors in different contexts
 \cite{G,SW,MN,M}. The
 Lax pair for PDE with quadratic nonlinearity 
 in Fourier form is analyzed in \cite{M} (compare with
 eqs.(\ref{lin21},\ref{lin22})).  

This work was supported by RFBR grants 03-01-06122 and 1716-2003-1

\end{document}